%% file: main.tex
\documentclass[conference]{IEEEtran}
\IEEEoverridecommandlockouts
\input{packages.tex}
\input{macros.tex}
\title{Distributing Quantum Circuits Using Teleportations }

\author{\IEEEauthorblockN{Ranjani G. Sundaram}
\IEEEauthorblockA{\textit{Department of Computer Science} \\
\textit{ Stony Brook University, NY}}
\and
\IEEEauthorblockN{Himanshu {Gupta}}
\IEEEauthorblockA{\textit{Department of Computer Science} \\
\textit{ Stony Brook University, NY}}
}

\begin{document}
\maketitle
\vspace*{-0.5in}

\begin{abstract}
Scalability is currently one of the most sought-after objectives in the field of quantum computing. Distributing a quantum circuit across a quantum network is one way to facilitate large computations using current quantum computers. In this paper, we consider the problem of distributing a quantum circuit across a network of heterogeneous quantum computers, while minimizing the number of teleportations (the communication cost) needed to implement gates spanning multiple computers. 
We design two algorithms for this problem.
 The first, called Local-Best, initially distributes the qubits across the network, then tries to teleport qubits only when necessary, with teleportations being influenced by gates in the near future. The second, called Zero-Stitching, \blue{divides the given circuit into sub-circuits such that each sub-circuit can be executed using zero teleportations and 
 the teleportation cost incurred at the borders of the sub-circuits is minimal.} We evaluate our algorithms over a wide range of 
 randomly-generated circuits \blue{as well as known benchmarks,} and compare their performance to prior 
 work. \blue{We observe that our techniques outperform the prior approach by a significant margin (up to $50$\%)}. 
\end{abstract}

\input{Introduction}
\input{Background}
\input{ProblemFormulation}

\input{ProposedTechniques}
\input{Evaluation}
\input{Conclusion.tex}

\bibliography{ref}
\end{document}

%% file: packages.tex
\usepackage{algorithm}
\usepackage{algorithmic}
\usepackage{amsfonts}
\usepackage{amsmath}
\usepackage{color}
\usepackage{esvect}
\usepackage{gensymb}
\usepackage{graphicx}
\usepackage{mathtools}
\usepackage[binary-units=true]{siunitx}
\usepackage{stfloats}
\usepackage{url}
\usepackage{xspace}
\usepackage{wrapfig}
\usepackage[normalem]{ulem}
\usepackage{enumitem}
\usepackage{subcaption}
\usepackage[sort,square,numbers]{natbib}
\DeclarePairedDelimiter{\ceil}{\lceil}{\rceil}
\makeatletter
\newcommand\fs@spaceruled{\def\@fs@cfont{\bfseries}\let\@fs@capt\floatc@ruled
  \def\@fs@pre{\vspace{0.5\baselineskip}\hrule height.8pt depth0pt \kern2pt}%
  \def\@fs@post{\kern2pt\hrule\relax\vspace{-0.4cm}}%
  \def\@fs@mid{\kern2pt\hrule\kern2pt}%
  \let\@fs@iftopcapt\iftrue}
\makeatother
\floatstyle{spaceruled}
\restylefloat{algorithm}

%% file: macros.tex
\newcommand{\eat}[1]{}

\newcommand{\red}[1]{{\color{black}#1}}

\newcommand{\blue}[1]{{\color{black}#1}}
\newcommand{\bleu}[1]{{\color{black}#1}}

\newcommand{\para}[1]{\smallskip \noindent {\bf #1}}
\newcommand{\softpara}[1]{\smallskip \noindent \underline{#1}}

\newcommand{\showOutline}{s}

\definecolor{darkgreen}{rgb}{0,0.5,0}
\definecolor{darkblue}{rgb}{0,0,0.5}

\newcommand{\outline}[2]{\if#1\showOutline {\color{darkgreen}#2} \fi}
\newcommand{\detailedoutline}[2]{\if#1\showOutline {\color{darkblue}#2} \fi}

\newcommand{\todo}[2]{\if#1\showOutline {#2} \fi}

\newcommand{\purp}[1]{} 

\def\blackbox{\hfill {\vrule height6pt width6pt depth0pt}}
\newcommand{\vsplem}{\vspace{0.025in}}
\newcommand{\vsp}{\vspace*{-0.1in}} 

\newtheorem{lem}{Lemma}
\newtheorem{thm}{Theorem}
\newtheorem{obv}{Observation}

\newenvironment{lem-wo-prf}{\begin{lem} \nopagebreak}{{\hfill$\blackbox$} \end{lem}}
\newenvironment{lem-w-prf}{\vsp \begin{lem} \nopagebreak}{\end{lem}}
\newenvironment{thm-w-prf}{\vsp \begin{thm} \nopagebreak}{\end{thm}}
\newenvironment{thm-wo-prf}{\begin{thm} \nopagebreak}{{\hfill$\blackbox$} \end{thm}}

\newcounter{packednmbr}

\newcommand{\eps}{\mbox{EP}\xspace}

\newcommand{\tabu}{{\tt Tabu}\xspace}

\newcommand{\lb}{\textsc{Local-Best}\xspace}
\newcommand{\zs}{\textsc{Zero-Stitching}\xspace}

\newcommand{\id}[1]{\textit{#1}}
\newcommand{\cz}{\id{CZ}\xspace} 

\newcommand{\dqc}{{\tt DQC}\xspace}
\newcommand{\zcsc}{{\tt ZCSC}\xspace}
\newcommand{\zcscs}{{\tt ZCSC}s\xspace}
\newcommand{\dqct}{{\tt DQC-T}\xspace}

\newcommand{\CNOT}{\ensuremath{\mathit{CNOT}}\xspace}
\newcommand{\CZ}{\ensuremath{\mathit{CZ}}\xspace}

\newtheorem{defin}{Definition}

%% file: Introduction.tex
\section{\bf Introduction}

Quantum computing is a recent method of computation that has the potential to solve some problems deemed intractable using classical computation. Quantum computers use the properties of quantum mechanics to analyze large numbers of possibilities and extract potential solutions to complex problems. However, the same properties raise challenges when we try to scale quantum computers to perform computations involving a large number of qubits. One approach to tackle these challenges is to distribute such computations across a quantum network.

We seek to design methods that distribute a given quantum program over a quantum network efficiently.
A natural optimization is to minimize the overall communication cost incurred in 
executing the distributed program over the given network. This problem is known to be NP-Hard, and has been considered recently in several works~\cite{andres2019automated,Daei2020OptimizedQC, g2021efficient}. There are two main means of viable quantum communication, viz., teleportations and cat-entanglements. 
Cat-entanglements, in essence, create read-only copies of a qubit; these copies
can thus be used as control operands to binary quantum gates. Though cat-entanglements facilitate higher distribution/concurrency, they have many disadvantages: 
(i) The 
copies need to be disentangled (i.e., 
withdrawn or destroyed) prior to any unary operation on the original qubit. 
(ii) Distribution algorithms based on cat-entanglements have assumed circuits composed of only binary CZ (due to their symmetry) and unary gates.
(iii) The shared copies, being entangled, incur a higher degree of decoherence. 
Thus, it is desirable to restrict communication modes to only teleportations.
In this work, we thus address the problem of distribution of quantum circuits using 
only teleportations.

\para{Our Contributions.} In this paper, we define the \dqct problem, which aims to distribute a given quantum circuit across a general quantum network with a minimum number of teleportations.
\dqct has been addressed before:~\cite{davarzani2020dynamic,Nikahd_2021, zomorodi2018optimizing} provide worst-case exponential-time algorithm under specialized settings, while~\cite{Daei2020OptimizedQC} provides a polynomial-time heuristic for the special case when the quantum computers in the network have identical memory capacities. In this paper, we design two novel polynomial-time algorithms for the general \dqc-t problem 
that outperform the algorithm in ~\cite{Daei2020OptimizedQC} by up to $50\%$.

%% file: Background.tex
\section{\bf Background}
\label{sec:Background}
\begin{figure}[t]
\vspace{-0.3in}
\centering
\includegraphics[width=0.35\textwidth]{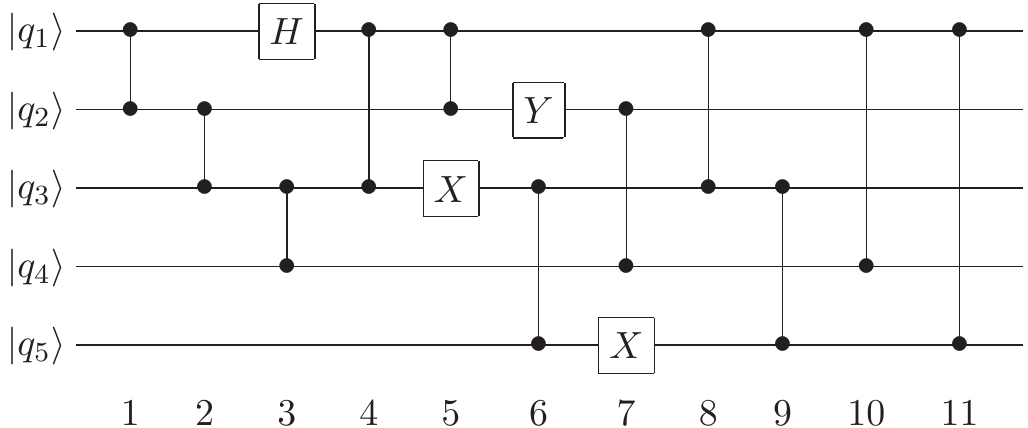}
\caption{Quantum Circuit Representation.}
\label{fig:bg-circuit-eg}
\end{figure}
We start with giving a brief background on two key quantum concepts relevant to our paper: quantum circuits and quantum 
communication methods.

\para{Quantum Circuits.}  Quantum computation is typically abstracted as a \emph{circuit}, where horizontal ``wires'' represent \emph{qubits} which carry quantum data, and operations on the qubits performed by vertical ``gates'' connecting the operand wires~\cite{nielsen_chuang_2010} (See Fig.~\ref{fig:bg-circuit-eg}).  Quantum computers (QCs) evaluate a circuit by applying the gates in the left-to-right order, so this circuit can also be understood as a sequence of machine-level instructions (gates) over a fixed number of data cells (qubits).

\softpara{Representation.}  Analogous to classical Boolean circuits, there are several universal gate sets for quantum computation: any quantum computation can be expressed by a circuit consisting only of gates from a universal gate set. We consider the universal gate set with unary and \CNOT (or \CZ) gates. We represent an \emph{abstract quantum circuit} $C$ over  a set of qubits $Q = \{q_1, q_2, \ldots\}$ as a sequence of gates $\langle g_1, g_2, \ldots \rangle$ where each $g_t$ is either binary \CNOT gate
or a unary gate. 
We thus represent binary gates in a circuit as triplets $(q_i, q_j, t)$ where $q_i$ and $q_j$ are the two operands, and $t$ is the time instant (see below) of the gate in the circuit, and unary gates as pairs $(q_i, t)$ where $q_i$ is the operand and $t$ is the time instant.
We use $N_q$ and $N_g$ to denote the number of qubits and gates in the circuit, 
respectively.

\begin{figure}[t]
\centering
\includegraphics[width=0.35\textwidth]{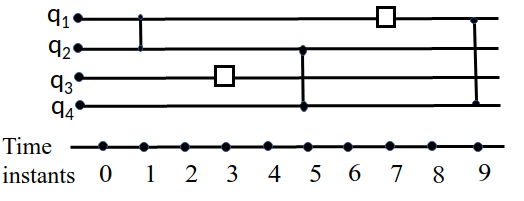}
\caption{The gates are at odd-numbered time instants $1, 3, \ldots, {9}$. The additional time instants introduced are the even-numbered time instants.}
\vspace*{-3ex}
\label{fig:time_instants}
\end{figure}

\softpara{Time Instants.} Each gate occurs uniquely at a time instant.
In addition to the instants where the gates occur, we introduce time instants in between consecutive gates for convenience; we enforce that teleportations occur only 
at these additional time-instants so that they don't interfere (or occur concurrently) 
with the gate operations. See Fig.~\ref{fig:time_instants}. 

\para{Quantum Communication.}
If a given quantum circuit is to be evaluated in a distributed fashion over a network of QCs, we have to first distribute the qubits over the QCs.  But such a distribution may induce gates in the circuit to span multiple QCs.  To execute such \emph{non-local} gates, we need to bring all operands' values into a single QC via quantum communication. 
However, direct/physical transmission of quantum data 
is subject to unrecoverable errors, 
as classical procedures such as amplified signals 
or re-transmission cannot be applied due to quantum no-cloning~\cite{wooterszurek-nocloning,Dieks-nocloning}.\footnote{Quantum error
correction mechanisms~\cite{muralidharan2016optimal,devitt2013quantum} can be used to mitigate the transmission errors, 
but their implementation is very challenging and is not expected to be used
until later generations of quantum networks.}
Fortunately, there are other viable ways to 
communicate qubits across network nodes, as described below.



\softpara{Teleportation.}
An alternative approach to physically transmit qubits is via \emph{teleportation}~\cite{Bennett+:93} 
which requires an a priori distribution of maximally-entangled pair (MEP) of qubits (e.g., Bell Pair) over the two nodes~\cite{tqe-22, pre-dist-22}.  With an MEP distributed over nodes $A$ and $B$,  teleportation of a qubit state from $A$ to $B$ can be accomplished using classical communication and local gate operations, while consuming/destroying the MEP. 

\blue{\softpara{Cat-Entanglement: Creating ``Linked Copies'' of a Qubit.}
Another means of communicating qubit states is by creating \emph{linked read-only copies} of a qubit across QCs, via \emph{cat-entanglement} operations~\cite{Eisert+:00,YimsiriwattanaL:05} which, like teleportation, require a Bell Pair to be shared \emph{a priori}. These linked copies are useful in efficient distributed evaluation of circuits involving only \cz and unary gates, as a single copy can be used as a control qubit for a contiguous sequence of \cz gates. These linked copies can only be used till there is a unary operation on the qubit.

However, restricting to circuits involving only \cz and unary gates means not exploiting potential circuit representations that contain fewer binary gates. Moreover, cat-entanglements entail the creation of entangled qubits that remain entangled for extended periods of time; these entangled qubits have a higher chance of decoherence and can lead to loss of fidelity of original qubits. Thus, in this work, we use only teleportations.}


%% file: ProblemFormulation.tex
\section{\bf Problem Formulation and Related Work}
\label{sec:ProblemFormulation}
We start by defining some terms and models before moving on to the problem formulation.

\para{Quantum  Network (QN).}
We represent a quantum network as a connected undirected graph with nodes representing QCs and edges representing  (quantum and classical) direct communication links. Nodes of the network are denoted by $P$; we use the words node, computer, and QC interchangeably.  We denote the number of computers/nodes in the network by $N_p$. Each node $p \in P$ has a storage capacity $s_p$ that represents the maximum number of qubits that can exist in node $p$ at any time instant. 

\para{Home-Computer (Function).}
To distribute the circuit, we first distribute the qubits of the circuit among the quantum computers. Later, our algorithms determine how to move these qubits around via teleportations, but, at any given time instant, each qubit is contained in a unique computer. Thus, at every time instant $t$, we can define a function $\pi^{t} :  Q \rightarrow P$ from qubits to the computers in the quantum network. This function specifies which computer a qubit is contained in at a time instant and is called the \emph{Home-Computer function}.  

Naturally, at any time instant, the home computer function must satisfy the storage capacity constraints of every computer. Therefore, for a computer $p$ and a time instant $t$, the number of qubits such that $\pi^t(q)=p$ must be less than $s_p$. 

\softpara{Non-Local Gates.}
A gate $(q_i, q_j, t)$ is considered to be {\em non-local} at time $t$ if $q_i$ and $q_j$ have different home-computers, i.e., $\pi^{t}(q_i) \neq \pi^{t}(q_j)$. Other gates are considered local gates.

\para{Communication (Teleportation) Cost.}
We represent a teleportation by a quadruplet $(q,p_1, p_2, t)$ which signifies that the qubit $q$ is teleported from computer $p_1$ to 
computer $p_2$ at time $t$. Note that $p_1$ must be equal to $\pi^{t-1}(q)$, and 
the teleportation $(q, p_1, p_2, t)$ results in $\pi^t(q)$ becoming $p_2$. 
As mentioned before, 
we restrict teleportations to only occur at non-gate time instants. 

The cost of a teleportation $(q, p_1, p_2, t)$ is defined as the distance between the nodes $p_1$  and $p_2$ in the quantum network graph, i.e., minimum number of hops between $p_1$
and $p_2$ in the network graph. 
This cost accounts for the fact that teleporting a qubit from $p_1$ to $p_2$ requires
an \eps over nodes $p_2$ and $p_1$ whose generation cost we assume to be proportional
to the distance between $p_1$ and $p_2$.  For simplicity, we assume that teleportation across two remote nodes does not use any storage space in the intermediate nodes. 

\para{\dqct Problem.} Given a quantum network and a quantum circuit, our objective is to find a valid home-computer function at each time instant of the circuit such that (i) all gate operations of the circuit are local, and (ii) the total teleportation cost incurred (due to changes in the home-computer function over time) is minimized.

Note that teleportations incurred can be uniquely determined from the home-computer function for every instant. The \dqct problem can be shown to be NP-Hard by a reduction from the graph bisection problem. We omit the proof here. 


\begin{figure}[t]
\centering
\includegraphics[width=0.35\textwidth]{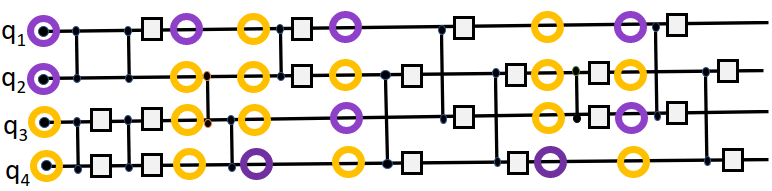}
\caption{Solution to a \dqct instance.}

\vspace*{-3ex}
\label{fig:teleExample}
\end{figure}

\para{Example 1.} Consider a \dqct problem instance in Figure~\ref{fig:teleExample}---a circuit with four qubits and two computers each with a storage memory of three qubits. We assume each pair of computers to be connected by a network link, and thus, the cost of any teleportation is one. A solution for this instance is also illustrated in the figure.  Initially, qubits $q_1$ and $q_2$ are assigned to the first computer (signified by purple circles), while $q_3$ and $q_4$ are assigned to the second computer (signified by yellow circles). As non-local gates are encountered, the home-computer function changes. Each time a particular qubit changes color, one teleportation occurs. The total cost of the solution provided is 5 (the number of teleportations).

\para{Related Work.}
Daei et al.~\cite{Daei2020OptimizedQC} consider the problem of reducing teleportation cost in distributed quantum circuits. The main idea behind their paper is to
construct an appropriate graph over qubits and use the Kernighan-Lin graph bisection algorithm iteratively to partition the graph into $k$ parts. Then, each part is assigned to a different quantum computer, and every non-local gate is
executed using teleportations. In another work, Moghadam et al.~\cite{zomorodi2018optimizing} consider the special case of only two computers with the qubits assignment across the computers
is already given; for this special case, they provide an exponential time algorithm that finds the minimum number of teleportations to execute all binary gates. Finally, Nikahd et al.~\cite{Nikahd_2021} first segregate the binary gates into ``levels,'' then determine the best partition of qubits for each level by solving an integer linear program. 

\blue{Some works have addressed the DQC problem using cat-entanglements as the only means of communication. For example,~\cite{andres2019automated} poses} the \dqc problem as a balanced hypergraph partitioning problem and gives a heuristic based on hypergraph min-cut. For a simplified setting,~\cite{g2021efficient} presents a two-step algorithm for the \dqc problem, wherein
the first step determines the partitioning of qubits to computers through balanced graph partitioning and the second step minimizes the number of cat-entanglement operations via
an iterative greedy approach. 
Finally, in ~\cite{dqc-gen}, the authors allow use of {\em both} 
teleportations and cat-entanglements to communicate quantum states, and
consider the general \dqc problem with both storage and execution memory constraints on computers. 
\blue{In this work, we focus on using only teleportations due to reasons mentioned in Section \ref{sec:Background}. }

%% file: ProposedTechniques.tex
\section{\bf \lb Algorithm}
\label{sec:LocalBest}

In this section, we describe our first algorithm, \textsc{Local-Best}.

\para{Basic Idea.}
The high-level idea behind the \lb algorithm is to teleport qubits only when necessary, with teleportations being influenced by gates in the near-future. \lb algorithm has two steps: (i) Finding an {\em initial} assignment of qubits to computers such that the number of resulting non-local binary gates is minimized, and (ii) \blue{For each non-local binary gate $g$, choose the teleportations to execute $g$ locally; the teleportations are chosen based on 
the "near future" such that the total number of teleportations is minimal.}

\floatname{algorithm}{Algorithm}
\begin{algorithm}[t]
\caption{\lb}
  \label{alg:LocalBest}  
  \begin{flushleft}
  \textbf{Input:} Quantum Network $N$, Quantum circuit $C$. \\
  \textbf{Output:} An initial home-computer function $\pi$ and a set of teleportations $\mathcal{T}$ that make all binary gates local.
  \end{flushleft}
  \begin{algorithmic}[1]
        \STATE $\pi \leftarrow$ \textsc{Tabu Search}($N,C$).
        \STATE $L_q\leftarrow$ preference order of computers depending on the next r binary gates on $q$.
        \STATE Uncovered $\leftarrow$ non-local binary gates in ${C}$ due to $\pi$.
        \FOR { $(q_1,q_2,t)$ in Uncovered }
            \STATE Update $L_{q_1}$ and $L_{q_2}$.
            \STATE $r1,r2=$ maximum index over the next $r$ binary gates on $q_1$ and $q_2$.
            \FOR{each computer $p$}
                \STATE $G_p \leftarrow $ graph over qubits in $p$ with edge weights $w(q_i,q_j)=$ number of gates of the form $(q_i,q_j,t')$ with $t'<=max(r_1,r_2)$.
                \STATE $W_1$ (and $W_2$) $=$ number of gates of the form $(q,\Bar{q}, t')$ where $q\in p$, $t'\leq r_1$ (and $t'\leq r_2$) and $\Bar{q}=q_1$ (and $q_2$).
                \STATE \label{algstep:tele} Pick one or two qubits in $p$ with minimum number of cut edges in $G_p$.
                \STATE $B_p=W_1+W_2-w(q_1,q_2)-r\times$(number of gaps needed in $p$)

            \ENDFOR
            \STATE Computer $p^* \leftarrow p$ with maximum $B_p$.
            \STATE $T_1\leftarrow$ teleportations that move qubits in step \ref{algstep:tele} to the highest available computers on their preference order. 
            \STATE $T_2 \leftarrow $ teleportations that move $q_1$, or $q_2$, or both to $p^*$.
            \STATE $\mathcal{T}\leftarrow \mathcal{T} \cup T_1 \cup T_2$.
        \ENDFOR
        \RETURN $\pi, \mathcal{T}$.
  \end{algorithmic}
\end{algorithm}


\para{Step 1. Initial Assignment of Qubits to Computers.} Similar to the approach in \cite{dqc-gen}, we assign qubits to computers using a Tabu Search~\cite{skorin1990tabu} heuristic,
in a way that satisfies the storage constraints at each node and minimizes the number of 
non-local gates.  Below, we define the three key aspects that characterize the 
Tabu Search: (i) a solution, (ii) a solution's neighbors, and 
(iii) the cost of a solution; here, we use a simpler  
cost model than that used in~\cite{dqc-gen}.

\softpara{Solution and Its Neighbors.}
In our context, a solution is a valid home-computer function.
Neighbors of a given solution $\pi$ can be defined as valid solutions $\pi'$ that result from
either: (i) changing the assignment/mapping of a single qubit without violating the storage
constraints, or (ii)  ``swapping'' of two qubits mapped to two different computers in $\pi$.

\softpara{Solution's Cost.}
A solution's cost  can be defined as the number of  non-local gates 
resulting from the solution's qubit assignment. 
More formally, the cost of a solution $\pi$, denoted by $\id{cost}(\pi)$ is
$$\id{cost}(\pi) = \sum_{q_1, q_2 \in Q} w(q_1, q_2) \times \text{distance}(\pi(q_1), \pi(q_2))$$
where $w(q_1, q_2)$ is the \emph{number} of  binary gates 
between $q_1$ and $q_2$ if they are assigned to different computers.

Based on the above concepts, we can do 
the initial assignment of qubits using Tabu Search
as follows.

\softpara{\tabu Algorithm.} 
\begin{enumerate}
\item $\pi^* = \pi =$ initial random solution
\item $L = [\ ]$ /* a bounded-length list of forbidden solutions */ 
\item Repeat for $\lambda$ iterations:
\begin{enumerate}
\item $\pi = \id{argmin}_{\pi' \in \id{neighbors}(\pi)-L}\ \id{cost}(\pi')$
\item $\pi^* = \pi$ if $\id{cost}(\pi) < \id{cost}(\pi^*)$
\item $L = L \cup \{\pi\}$, removing the oldest element from $L$ if necessary to maintain length bound.
\end{enumerate}
\item Return $\pi^*$
\end{enumerate}

\para{Step 2. Determining Teleportations.} 
We now describe a procedure that determines a small number of
teleportations to execute the non-local binary gates. 
The high-level idea of our approach is as follows:

\begin{itemize}
    \item Scan the circuit from left to right and determine teleportations to execute each non-local  gate encountered. Here, a gate is considered non-local or local based on the
    home-function at {\em that} time instant.
    \item 
    For a non-local binary gate $g$ encountered, determine the ``best'' computer where 
    $g$ should be executed by teleporting its operands to. Here, the best computer is determined (see below) based on the next (following) $r$ binary gates that share an operand with $g$.
\end{itemize}

\softpara{Best Computer to Execute $g$.}
More formally, to determine the best computer for a non-local gate $g$, we define and compute 
a ``benefit'' for each computer and pick the computer with the highest benefit. 
For a non-local binary gate $g = (q_1,q_2)$, the benefit of a computer $p$ is define as 
the number of gates among the next $r$ binary gates involving $q_1$ and/or $q_2$ 
that become local (from non-local) by teleporting $q_1$ and $q_2$ to $p$. 
We note that in order to execute $(q_1,q_2)$ in computer $p$, 
both $q_1$ and $q_2$ must be in $p$, and
teleporting $q_1$ and/or $q_2$ would require one or two free storage spaces 
in $p$. 
If needed, the desired storage space(s) in $p$ can be created by 
teleporting some other qubits from $p$ to another computer.
\blue{For simplicity, such supplementary teleportations are not taken into account in computing $p$'s benefit.}

The overall \lb algorithm is presented in Algorithm~\ref{alg:LocalBest}.




\section{\bf \zs Algorithm}
\label{sec:ZeroStitch}

\begin{figure}[t]
\centering
\begin{subfigure}{0.33\textwidth}
\includegraphics[width=\textwidth]{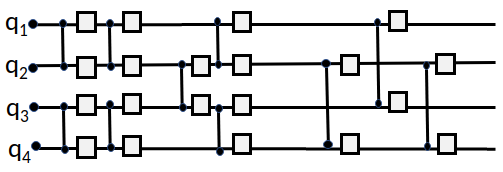}
\vspace{-0.5cm}
\caption{}
\vspace*{0.2in}
\label{fig:splitA}
\end{subfigure}
\begin{subfigure}{0.36\textwidth}
\includegraphics[width=\textwidth]{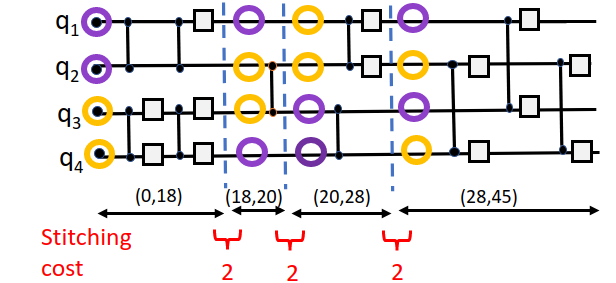}
\caption{}
\label{fig:splitB}  
\end{subfigure}
\caption{Division of a circuit into four zero-cost sub-circuits $(0,18),(18,20),(20,28)$, and $(28,45)$, and stitching costs for the adjacent pairs of zero-cost sub-circuits. }
\label{fig:split}
\vspace{-0.5cm}
\end{figure}

We now describe the \zs algorithm which consists of two high-level steps: (i) identifying ``zero-cost'' sub-circuits, i.e., contiguous sub-circuits that can be executed without any
teleportations; (ii) dividing the given circuit into zero-cost sub-circuits 
and ``stitching'' them together using teleportations. 
We start with defining a few concepts that will 
help in formalizing the overall \zs algorithm. 

\para{Sub-Circuit.}
A sub-circuit of a given circuit $C$ is any contiguous part of $C$, and can be represented by the starting and the ending time-instant. For example, in Figure~\ref{fig:splitB}, $(0,18),(18,20),(20,28)$, and $(28,45)$ are some of the sub-circuits.

\para{Zero-cost Sub-circuit (\zcsc).} 
 We \red{deem} a sub-circuit $C$ to be {\em zero-cost} if,
 for \red{{\em some} fixed} home-computer function $\pi$,
 all gates in $C$ become local; here, by {\em fixed} we mean that the home-computer function $\pi$ remains unchanged throughout the sub-circuit $C$. We use the acronym \zcsc for zero-cost sub-circuit.  
 In figure~\ref{fig:splitB}, the given circuit has been divided into four \zcscs: $(0,18),(18,20),(20,28)$, and $(28,45)$, with two network nodes.  Here, for each of the four \zcscs, the figure
 also gives the home-computer function (represented by yellow and purple circles) which makes all the gates local.

\para{Stitching Two Sub-circuits.} Consider 
two (adjacent) \zcscs $C_{1}$ and $C_{2}$, and the corresponding home-computer functions $\pi_{1}$ and $\pi_{2}$ that make all their gates local.
Then, stitching $C_{1}$ to $C_{2}$, for the
given home-computer functions $\pi_{1}$ and $\pi_{2}$,
signifies determining and using a set of teleportations that transform $\pi_{1}$ to $\pi_{2}$;
the stitching cost is considered to be equal to the teleportation cost.

For example, in figure \ref{fig:splitB}, stitching the \zcsc $(0,18)$ to $(18,20)$ incurs two teleportations: teleporting $q_2$ to the yellow computer and $q_4$ to the purple computer. Note that stitching of minimum cost for a given pair of \zcscs and corresponding home-computer functions is trivial to determine. However, if we allow ``relabelling'' of computers, determining the minimum-cost stitching requires computing the maximum-weight matching in an appropriate bipartite graph (see the SCB() function below).

Using the above concepts, we now describe the two steps of \zs algorithm.

\para{Step 1. Identifying Zero-Cost Sub-circuits.} In this step, we consider all possible sub-circuits of $C$ and, for each sub-circuit $C_i$, we try to determine a home-computer function that makes all gates of $C_i$ local. Since determining such a home-computer function is NP-Hard, we use a heuristic based on bin-packing approximation algorithm as described below.

For every sub-circuit $C_i$ of given circuit $C$, we do the following:
    \begin{itemize}
        \item Construct a graph $G_{C_{i}}$ with the given circuit's qubits as vertices, and an edge between two qubits if there are gates between the two qubits in the sub-circuit $C_i$. 
        \item Consider the connected components of $G_{C_i}$. 
        \item ``Pack'' the connected components into the given quantum computers, such that the qubits of any connected component are all within a single computer and the total number of qubits in any computer $p$ is less than $p$'s storage capacity. If we can find such a packing, then $C_i$ is considered a {\em zero-cost sub-circuit}. Note that determining such a packing is equivalent to solving the well-studied {\em bin-packing problem} with computers as bins of size equal to their storage capacities and connected components of $G_{C_i}$ as items of size equal to the number of qubits in the connected component. In our work, we use the first-fit greedy algorithm~\cite{Dsa2013FirstFB} to determine the packing of connected components into computers. 
        
    \end{itemize}

The above procedure determines a set of \zcscs. For each \zcsc $C_i$, the packing algorithm also gives the \red{(fixed)} home-computer function $\pi_i$ 
which makes all gates local;  we
refer to $\pi_i$ as the \red{{\em pre-computed}} home-function for $C_i$.

\para{Step 2. Efficient Division of $C$ into Zero-Cost Sub-Circuits.} We use dynamic programming to find a \red{division} of $C$ into a sequence of \zcscs and their corresponding home-computer functions such that the total cost of stitching adjacent sub-circuits is minimal. We first present a simpler version of the dynamic programming (DP) procedure, and then improve 
it further.

\para{\textsc{Simple DP}:} We can divide a given circuit $C$ into \zcscs with minimum aggregate stitching cost using dynamic programming as follows.
Let $S[1, i, j]$ be the optimal cost for the sub-circuit $(1,j)$, formed by stitching together
multiple \zcscs within $(1,j)$ with the constraint that $(i,j)$ is the last such
\zcsc. Our goal is to determine $\min_i S[1, i, m]$, where $m$ is the last time instant. We determine $S[\ ]$ values recursively using dynamic programming as follows. 
\begin{align} \label{eqn:simpledp}
    S[1, i, j]=    min_{i’} \ S[1, i’, i]  + \textsc{SC}((i',i),(i,j))
\end{align}

Above, $\textsc{SC}((i',i),(i,j))$ is a function that 
gives the cost of stitching the 
\zcsc $(i',i)$ to $(i,j)$ for the given pre-computed home-computer
functions. As mentioned above,  
the function $\textsc{SC}()$ is straightforward if we do not allow relabelling of computers.
To allow relabelling of computers, we define a more general stitching function 
called $\textsc{SCB}()$ that uses a maximum matching algorithm. Note that
relabelling of computers is possible only when the computers have 
uniform storage 
capacity. 
We describe $\textsc{SCB}()$ below. 

\softpara{$\textsc{SCB}$($C_1,C_2$).}
    \begin{enumerate}
    \item \label{scb1} Let $\pi_{C_1}$ and $\pi_{C_2}$ be the given pre-computed home-computer functions associated with $C_1$ and $C_2$ respectively.
    \item \label{scb2} Denote by $P$ and $\Bar{P}$ the partition of qubits corresponding to $\pi_1$ and $\pi_2$ respectively. Here, each element of $P$ or $\Bar{P}$ is a set of qubits mapped to a particular network node.
    \item \label{scb3} Construct a bipartite graph over elements of $P$ and $\Bar{P}$ as vertices. For each pair of elements $P_i$ and $\Bar{P_j}$ from $P$ and $\Bar{P}$ respectively, we associate an edge weight of $w(P_i, \Bar{P_j})$ equal to the number of qubits contained in both $P_i$ and $\Bar{P_j}$.
    \item \label{scb4} Find a maximum-weight matching $M$ in the above graph.
    \item \label{scb5} Determine teleportations corresponding to the edges in the matching $M$.
    \end{enumerate}
We now motivate and then discuss an improved version of the above \textsc{Simple-DP} algorithm that recomputes the home-computer function for sub-circuit $(i,j)$ in Eqn.~\ref{eqn:simpledp}.



\softpara{Motivation for Further Improvement.} 
The above \textsc{Simple-DP} algorithm 
yields a solution to the \dqct problem by 
computing $min_{i'} S[1,i',m]$, where $m$ is the number of binary gates.
\bleu{In fact,} it is an optimal way to
divide the input circuit into \zcscs {\em given the pre-computed} 
home-functions for the \zcscs. However, it is possible
that a different set of home-functions for the \zcscs may yield
a better (i.e., with lower total stitching cost) \dqct solution. 
Note that the pre-computed home-functions are 
computed independently for each \zcsc, 
while, for an optimal \dqct solution, the
home-computer functions of the \zcscs being 
stitched should be as ``close'' to each other as possible.
To incorporate the above insight, we modify the above \textsc{Simple-DP} 
algorithm as below.

\para{\textsc{Improved DP}.}
Consider Eqn.~\ref{eqn:simpledp}.
In the \textsc{Improved-DP} algorithm, 
we allow re-computation of 
the home-computer function  corresponding to 
the \zcsc $(i,j)$ 
based on the home-computer function of the preceding
\zcsc $(i',i)$, when computing the optimal cost $S[1,i,j]$ for $(1,j)$. 
This modification is incorporated into the improved stitching function $\textsc{SC}^*$, as described \bleu{momentarily}.

As before, let $S[1, i, j]$ be the optimal cost for the sub-circuit $(1,j)$, formed by stitching together
multiple \zcscs within $(1,j)$ with the constraint that $(i,j)$ is the last such
\zcsc. Our goal is to determine $\min_i S[1, i, m]$, where $m$ is the last time instant. The improved dynamic programming formulation to determine $S[\ ]$ values recursively 
is as follows. 
\begin{align}
\label{eqn:improveddp}
S[1, i, j] =    min_{i’} \ S[1, i’, i] + {\textsc{SC}^*}(\pi_{(i',i)},(i,j)).
 \end{align}
Above, $\pi_{(i',i)}$ is the home-computer function of the sub-circuit $(i',i)$ in 
the solution $S[1, i', i]$.\footnote{\bleu{In function $\textsc{SC}$, the home-computer functions
for the operands were implicit (i.e., the pre-computed home-computer functions), while 
in $\textsc{SC}^*$ we need to explicitly pass the home-computer function of 
$(i',i)$, based on which the home-computer function of 
$(i,j)$ is recomputed.
This is facilitated by storing two values associated 
with a solution $S[1,i,j]$---the cost of $(1,j)$ and 
the home-computer function for $(i,j)$; 
the latter is computed within the $\textsc{SC}^*$ function for the best $i'$ in Eqn.~\ref{eqn:improveddp}.}}
The improved stitching cost function $\textsc{SC}^*$ works in two steps: (i) For the input operand $(i,j)$ sub-circuit, 
it computes a home-computer function 
that ``matches'' as closely as possible with the input $\pi_{(i',i)}$---so that the cost of stitching the sub-circuit $(i', i)$
to $(i,j)$ is minimized, and (ii) it determines the teleportations  
required to stitch the sub-circuit $(i', i)$
to $(i,j)$ using the above home-computer functions. 


The final improved stitching function $\textsc{SC}^*()$ is as follows.

\softpara{$\textsc{SC}^*$($\pi_{C_1},C_2$):}
    \begin{enumerate}
    \item Let $\pi_{C_1}$ be the home-computer function associated with $C_1$ (given as input).
    \item \label{sc*2} Compute $\pi_2$ for $C_2$ (as described below).
    \item Compute teleportations that transform $\pi_{C_1}$ to $\pi_{C_2}$.
    \end{enumerate}
If we do not allow relabelling of computers, then the last step is straightforward. If we do allow relabelling (which is only possible for balanced nodes), we replace the last step of $\textsc{SC}^*$ with 
Steps~\ref{scb2}-\ref{scb5} of $\textsc{SCB}$; we call the resulting algorithm $\textsc{SCB}^*$.

\softpara{Computing $\pi_2$ for $C_2$.} We compute $\pi_2$ for $C_2$ in the second step of $\textsc{SC}^*$ as follows. \bleu{Essentially, we seek to compute a home-computer function $\pi_2$ for $C_2$ based on the home-computer function $\pi_1$ such that (i) $\pi_2$ makes all the gates of $C_2$ local, and (ii) the cost of stitching $\pi_1$ to $\pi_2$ is minimized. 
Consider an undirected graph $G_{c2}$ over qubits in $C_2$ wherein there is an edge $(q_i, q_j)$ if and only if there is a binary gate between $q_i$ and $q_j$.
First, to satisfy the first condition above (i.e., to make all gates of $C_2$ local), all the qubits in each connected component of $G_{c2}$ must entirely lie in one computer/node. 
Now, to minimize the cost of stitching $\pi_1$ to $\pi_2$, 
we consider a weighted graph $G'_{c2}$ 
over the connected components of $G_{c2}$ as vertices, wherein the weight between two connected components $cc_1$ and $cc_2$ signify the cost of keeping them in two different network nodes/computers.\footnote{\red{The weight between two connected components $cc_1$ and $cc_2$ is set to be the number of pairs of qubits $(q_1,q_2)$ such that $q_1$ is in $cc_1$, $q_2$ is in $cc_2$ and $\pi(q_1)=\pi_1(q_2)$.}}
Then, we find an assignment of connected components of $C_2$ (i.e., vertices of $G'_{c2}$) to the network nodes, by
partitioning the vertices into $k$ sets
(where $k$ is the number of network nodes) such that the cut over these partitions
is minimized and the number of qubits mapped to $i^{th}$ set is less than the storage
capacity of $i^{th}$ computer. 
We find such an assignment using a generalized version of the Tabu search algorithm described in Section~\ref{sec:LocalBest}.}

\floatname{algorithm}{Algorithm}
\begin{algorithm}[t]
\caption{\zs}
  \label{alg:ZeroStitch}  
  \begin{flushleft}
  \textbf{Input:} Quantum Network $N$, Quantum circuit $C$. \\
  \textbf{Output:} An initial home-computer function $\pi$ and a set of teleportations $\mathcal{T}$ that make all binary gates local.
  \end{flushleft}
  \begin{algorithmic}[1]
        \STATE $\mathcal{C} \leftarrow$ set of all sub-circuits of $C$.
        \FOR { each sub-circuit $S \in \mathcal{C}$  }
            \STATE $G_S \leftarrow$ a graph where the qubits form vertices and edges between qubits that have gates between them in $S$.
            \STATE $\pi_{S}\leftarrow$ a home-computer function such that all gates in $S$ are local (if it does not exist, $\pi_S=\emptyset$).
        \ENDFOR 
        \STATE $S_1,\dots, S_k\leftarrow$ the best sequence of sub-circuits obtained by running \textsc{Dynamic Programming} over $\mathcal{C}$.
        \STATE $T_1,\dots T_{k-1} \leftarrow$ teleportations obtained by stitching $\pi_{S_i}$ to $\pi_{S_{i+1}}$.
        \STATE $\mathcal{T}=T_1\cup \dots \cup T_{k-1}$.
        \RETURN $\pi_{S_1}, \mathcal{T}$.
  \end{algorithmic}
\end{algorithm}

%% file: Evaluation.tex
\section{\bf Evaluation}

\begin{figure*}[h]
    \centering
    \includegraphics[width=0.4\textwidth]{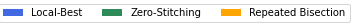}
    \vspace{-0.5cm}
\end{figure*}
\begin{figure*}[t]
\centering
\begin{subfigure}[b]{0.32\linewidth}
\includegraphics[width=\textwidth]{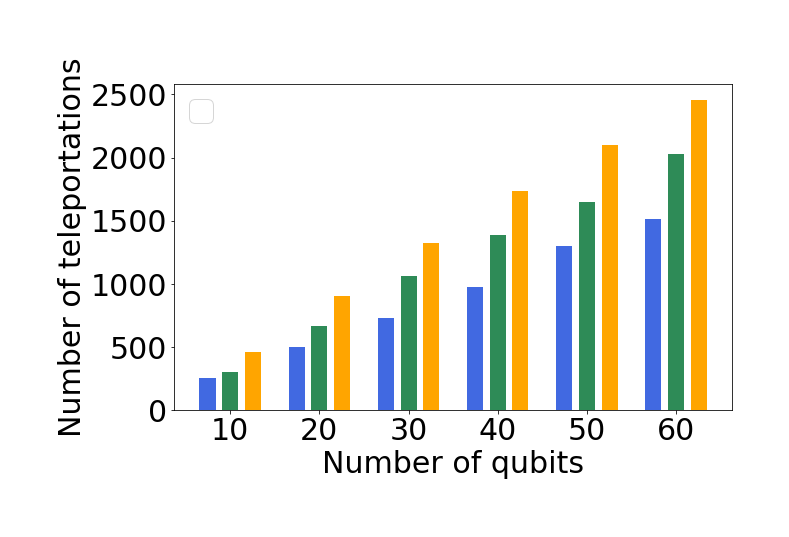}
\vspace{-0.6cm}
\caption{}
\label{fig:VaryQubits}
\end{subfigure}%
\begin{subfigure}[b]{0.32\linewidth}
\includegraphics[width=\textwidth]{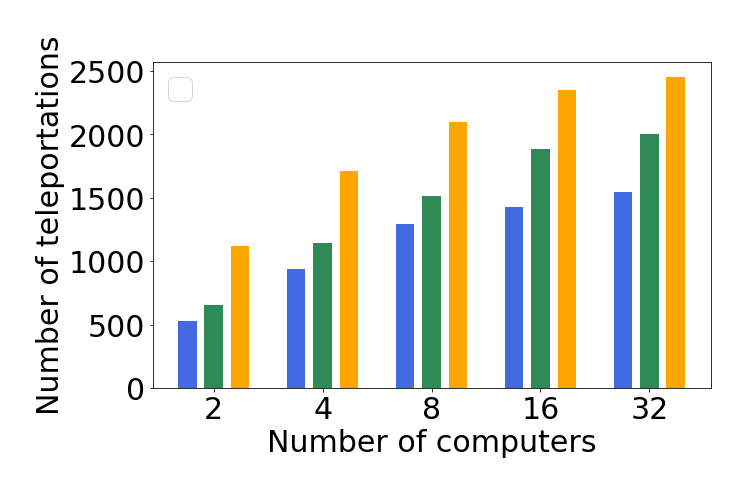}
\vspace{-0.6cm}
\caption{}
\label{fig:VaryPart}
\end{subfigure}%
\begin{subfigure}[b]{0.32\linewidth}
\includegraphics[width=\textwidth]{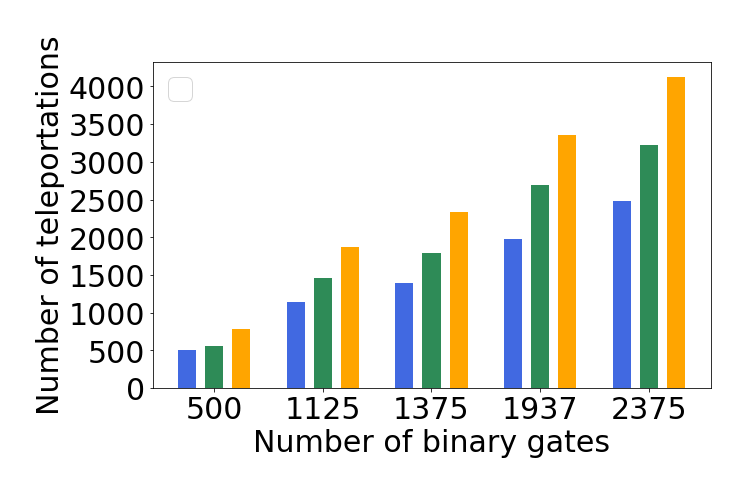}
\caption{}
\label{fig:VaryFracCZ}
\end{subfigure}
\begin{subfigure}{0.32\linewidth}
\includegraphics[width=\textwidth]{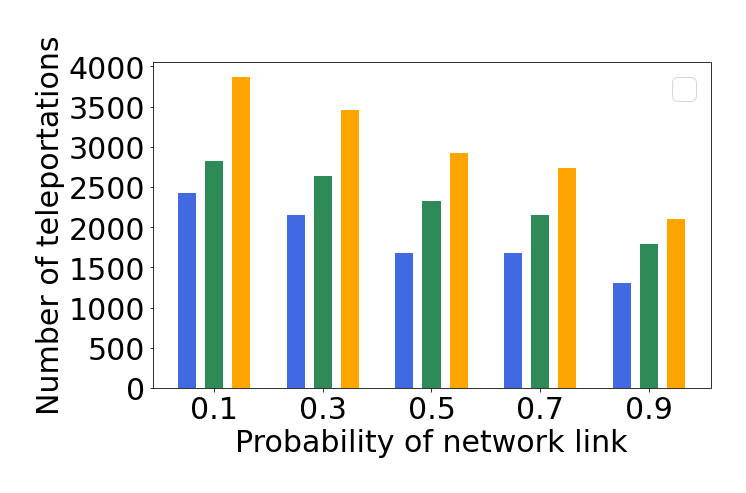}
\vspace{-0.7cm}
\caption{}
\label{fig:VaryProbability}
\end{subfigure}%
\begin{subfigure}{0.32\linewidth}
\includegraphics[width=\textwidth]{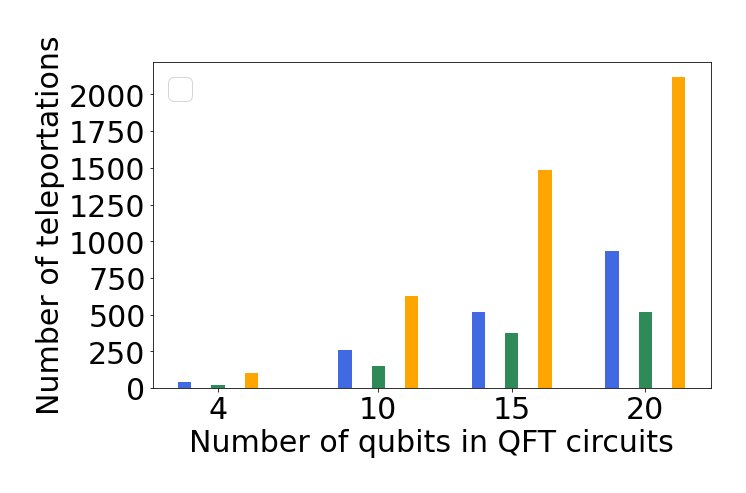}
\vspace{-0.7cm}
\caption{}
\label{fig:qft}
\end{subfigure}%
\begin{subfigure}{0.32\linewidth}
\includegraphics[width=\textwidth]{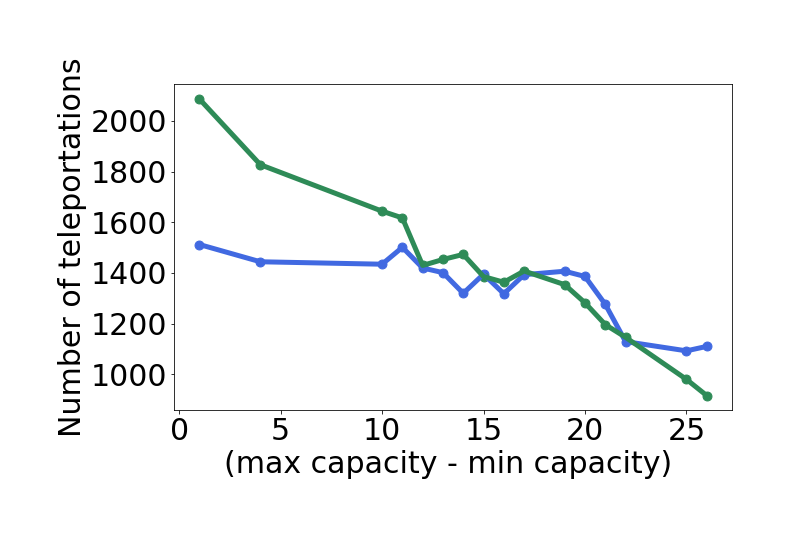}
\caption{}
\label{fig:VaryUnbalanced}
\end{subfigure}
\caption{Total communication cost incurred by different algorithms for varying parameter values.}
\end{figure*}
In this section, we \blue{evaluate our algorithms over both randomly generated and some benchmark quantum circuits.} We use the overall communication cost (i.e., the number of teleportations used) as the performance metric for our evaluations. We compare our 
techniques with prior work.

\para{Algorithms Compared.} We primarily compare the performance of three algorithms-(i) \lb from Section \ref{sec:LocalBest}, (ii) \zs from Section \ref{sec:ZeroStitch} \red{(which uses $\textsc{SCB}^*$)}, and (iii) \textsc{Repeated Bisection} from \cite{Daei2020OptimizedQC} which uses the Kernighan-Lin graph-bisection algorithm iteratively and implements non-local binary gates using two teleportations each. The other prior works on the \dqct problem have proposed algorithms that
are exponential-time in the worst case, and hence, not compared. \blue{In particular, Nikahd et al.~\cite{Nikahd_2021} use an integer linear program within their algorithm which takes a prohibitively long time to run even using ILP-solvers~\cite{cvxpy} on moderately sized inputs.}

\para{Generating Random Quantum Circuits.} We generate random quantum circuits using the following set of parameters:
\begin{itemize}
    \item Varying number of qubits (default value=$50$)
    \item Varying number of gates per qubit (default value =$50$)
    \item Varying fraction $f$ of binary gates (default value=$0.5$)
\end{itemize}
For a given set of parameter values, we generate the random input circuit sequentially, and at each point determine whether the next gate is binary with probability $f$. Operands of a gate are chosen randomly. 

\para{Generating Random Quantum Networks.} We generate a quantum network using the following set of parameters (as in~\cite{dqc-gen, g2021efficient}).
\begin{itemize}
    \item Number of computers (default value=$8$)
    \item Probability of a link between a pair of nodes (default=$1$)
\end{itemize}
We use the Python-based library~\cite{networkx} to generate connected  Erdős-Rényi graphs with a given edge probability. We assign values to number of vertices and probability of an edge in a way that the resulting Erdős-Rényi graph is connected with high probability. For most of the graphs used in our evaluation, we assume equal qubit storage capacities for all the nodes. This is because \textsc{Repeated Bisection} algorithm, with which our techniques are compared, is restricted to a uniform storage setting. 
However, we note that our techniques are general, and work for non-uniform storage capacities. In Fig.~\ref{fig:VaryQubits}-\ref{fig:qft}, we assume that the capacity of each computer is exactly $\ceil*{\frac{n}{N_p}}+1$ where $n$ is the number of qubits and $N_p$ is the number of computers.

\para{Evaluation Results.} We evaluate the algorithms by running them on randomly generated circuits and networks as described above. We vary one parameter at a time, keeping the others fixed to their default values. See Fig.~\ref{fig:VaryQubits}-\ref{fig:VaryUnbalanced}. We observe the following. 
\begin{itemize}
\item \lb performs best, followed by \zs, and then \textsc{Repeated Bisection}. \zs offers a $30\%$ reduction in cost compared to \textsc{Repeated Bisection} while \lb offers almost $50\%$ reduction in cost. 
\item We see that \zs performs worse than \lb on all randomly generated inputs presented here; we believe this is due to choosing gates uniformly at random. For benchmark circuits with repeated patterns, \zs performs better (see Fig.~\ref{fig:qft}). In addition, since all of the cost in \zs is in the form of stitching costs, 
we believe that \zs's performance can be improved with a more \red{creative}  algorithm for recomputing the home-computer function $\pi_2$ for $C_2$ in the second step of \textsc{$SC^*$}. 
\end{itemize}
We also note the following. In Fig.~\ref{fig:VaryQubits}, the cost of all algorithms steadily increase with increasing number of qubits. In Fig.~\ref{fig:VaryPart}, we see that the cost of \lb starts to plateau. In Fig.~\ref{fig:VaryFracCZ}, we see that the cost of 
each algorithm steadily increases with an increase in the number of binary gates. In Fig.~\ref{fig:VaryProbability}, the total cost decreases with an increase in the network link probability; this is because, in denser graphs, the teleportation cost between a pair of nodes is lower due to shorter distance between the nodes. 

In Fig.~\ref{fig:VaryUnbalanced}, we measure the performance of \lb and \zs on networks with nodes with non-uniform storage 
capacities. Here, the $x$-axis is the difference  between the maximum and minimum storage
capacity in the network. Note that \textsc{Repeated Bisection} algorithm is not shown
in this graph as it does not support such networks. \red{Here, ZC algorithm performs better than local-best for networks with high
imbalance in storage capacities.}

%% file: Conclusion.tex
\section{\bf Conclusion}

In this paper, we consider the problem of distributing quantum circuits over a quantum network while minimizing
the total teleportation cost. In our future work, we would like to further improve the performance of 
\zs by improving the stitching step. In addition, we would like to extend our approaches to $n$-ary gates.



%% file: main.bbl
\begin{thebibliography}{10}

\bibitem{cvxpy}
Cvxpy.
\newblock \url{https://www.cvxpy.org/}.

\bibitem{networkx}
Networkx.
\newblock
  \url{https://networkx.org/documentation/stable/reference/generated/networkx.generators.random_graphs.erdos_renyi_graph.html}.

\bibitem{andres2019automated}
P.~A.~Martinez and C.~Heunen.
\newblock Automated distribution of quantum circuits via hypergraph
  partitioning.
\newblock {\em Physical Review A}, 2019.

\bibitem{Bennett+:93}
C.~H. Bennett, G.~Brassard, C.~Crépeau, R.~Jozsa, A.~Peres, and W.~K.
  Wootters.
\newblock Teleporting an unknown quantum state via dual classical and
  {Einstein–Podolsky–Rosen} channels.
\newblock {\em Phys. Rev. Lett.}, 70(13), 1993.

\bibitem{Daei2020OptimizedQC}
O.~Daei, K.~Navi, and M.~Z. Moghadam.
\newblock Optimized quantum circuit partitioning.
\newblock {\em ArXiv}, abs/2005.11614, 2020.

\bibitem{davarzani2020dynamic}
Z.~Davarzani, M.~Z.~Moghadam, M.~Houshmand, and M.~Nouri-baygi.
\newblock A dynamic programming approach for distributing quantum circuits by
  bipartite graphs.
\newblock {\em Quantum Information Processing}, 19(10):1--18, 2020.

\bibitem{devitt2013quantum}
S.~J Devitt, W.~J Munro, and K.~Nemoto.
\newblock Quantum error correction for beginners.
\newblock {\em Reports on Progress in Physics}, 76(7):076001, 2013.

\bibitem{Dieks-nocloning}
D.~{Dieks}.
\newblock {Communication by {EPR} devices}.
\newblock {\em Physics Letters A}, 1982.

\bibitem{Dsa2013FirstFB}
G.~D{\'o}sa and J.~Sgall.
\newblock First fit bin packing: A tight analysis.
\newblock In {\em STACS}, 2013.

\bibitem{Eisert+:00}
J.~Eisert, K.~Jacobs, P.~Papadopoulos, and M.B. Plenio.
\newblock Optimal local implementation of non-local quantum gates.
\newblock {\em Phys. Rev. A}, 2000.

\bibitem{g2021efficient}
R.~G.~Sundaram, H.~Gupta, and CR~Ramakrishnan.
\newblock Efficient distribution of quantum circuits.
\newblock In {\em DISC}, 2021.

\bibitem{pre-dist-22}
M.~Ghaderibaneh, H.~Gupta, C.R. Ramakrishnan, and E.~Luo.
\newblock Pre-distribution of entanglements in quantum networks.
\newblock In {\em IEEE QCE}, 2022.

\bibitem{tqe-22}
M.~Ghaderibaneh, C.~Zhan, H.~Gupta, and C.~R. Ramakrishnan.
\newblock Efficient quantum network communication using optimized
  entanglement-swapping trees.
\newblock {\em IEEE Transactions on Quantum Engineering}, 2022.

\bibitem{muralidharan2016optimal}
S.~Muralidharan, L.~Li, J.~Kim, N.~L{\"u}tkenhaus, M.~D. Lukin, and L.~Jiang.
\newblock Optimal architectures for long distance quantum communication.
\newblock {\em Scientific reports}, 6(1):1--10, 2016.

\bibitem{nielsen_chuang_2010}
M.~A. Nielsen and I.~L. Chuang.
\newblock {\em Quantum Computation and Quantum Information}.
\newblock Cambridge University Press, 2010.

\bibitem{Nikahd_2021}
E.~Nikahd, N.~Mohammadzadeh, M.~Sedighi, and M.~S. Zamani.
\newblock Automated window-based partitioning of quantum circuits.
\newblock {\em Physica Scripta}, 2021.

\bibitem{skorin1990tabu}
J.~Skorin-Kapov.
\newblock Tabu search applied to the quadratic assignment problem.
\newblock {\em ORSA Journal on computing}, 2(1):33--45, 1990.

\bibitem{dqc-gen}
R.~G. Sundaram, H.~Gupta, and C.~R. Ramakrishnan.
\newblock Distribution of quantum circuits over general quantum networks.
\newblock In {\em IEEE QCE}, 2022.

\bibitem{wooterszurek-nocloning}
W.~K. {Wootters} and W.~H. {Zurek}.
\newblock {A single quantum cannot be cloned}.
\newblock {\em Nature}, 299(5886), October 1982.

\bibitem{YimsiriwattanaL:05}
A.~Yimsiriwattana and S.~J. {Lomonaco Jr.}
\newblock Generalized {GHZ} states and distributed quantum computing.
\newblock {\em AMS Cont. Math.}, 381(131), 2005.

\bibitem{zomorodi2018optimizing}
M.~Z.~Moghadam, M.~Houshmand, and M.~Houshmand.
\newblock Optimizing teleportation cost in distributed quantum circuits.
\newblock {\em International Journal of Theoretical Physics}, 57(3):848--861,
  2018.

\end{thebibliography}
